\documentclass[a4paper]{jpconf}
\usepackage{graphicx}
\usepackage{amsmath}
\usepackage{wrapfig}

\newcommand{\boutxx}{\texttt{BOUT++}}

\begin{document}
\title{Towards nonaxisymmetry; initial results using the Flux Coordinate Independent method in \boutxx}

\author{B W Shanahan, P Hill, B D Dudson}

\address{York Plasma Institute, Department of Physics, University of York, Heslington, York YO10 5DD, UK}

\ead{bws502@york.ac.uk}

\begin{abstract}
Fluid simulation of stellarator edge transport is difficult due to the complexities of mesh generation; the stochastic edge and strong nonaxisymmetry inhibit the use of field aligned coordinate systems.  The recent implementation of the Flux Coordinate Independent method for calculating parallel derivatives in \boutxx~has allowed for more complex geometries.  Here we present initial results of nonaxisymmetric diffusion modelling as a step towards stellarator turbulence modelling.  We then present initial (non-turbulent) transport modelling using the FCI method and compare the results with analytical calculations.  The prospects for future stellarator transport and turbulence modelling are discussed.  

\end{abstract}



\section{Stellarator and Nonaxisymmetric Modelling}
\label{sec:stellaratorintro}
Historically, neoclassical transport has been the dominant loss mechanism in stellarators~\cite{Ho1987}.  Recent optimizations have allowed for the minimization of neoclassical losses, which has culminated in the design of the Wendelstein 7-X stellarator~\cite{Grieger1992, Nuhrenberg1995}.  As Wendelstein 7-X has been optimized for neoclassical transport, turbulent transport could potentially become comparable to neoclassical losses.  As such, it is becoming increasingly important to simulate turbulence in nonaxisymmetric configurations to determine its role in comparison to neoclassical transport in order to optimize performance.  

In the core of stellarators, the closed flux surfaces and low collisionality facilitate the use of gyrokinetic codes such as GENE~\cite{Gorler2011}, which is currently the only technique for simulating stellarator turbulence.  However, due to the small scales simulated in gyrokinetics, the computation is quite expensive for experimentally relevant temporal and spatial scales.  Additionally, GENE simulations are currently localized to single flux surfaces or flux-tube geometries.

The edge of stellarators, however, includes stochastic regions and magnetic islands, and edge modelling in stellarators is currently limited to magnetohydrodynamic transport modelling to determine the steady state profiles.  This is done primarily using EMC3~\cite{Feng1999}, which employs a Monte-Carlo solver to simulate three dimensional transport (non-turbulent) equations in order to determine steady state quantities for divertor profiles. The relatively high collisionality of edge plasmas both in stellarators and tokamaks justifies a fluid approach to turbulence simulations, however the current nature of plasma fluid turbulence simulations renders it difficult to simulate nonaxisymmetric magnetic geometries; stochastic regions are difficult in field-aligned coordinate systems.

Edge fluid turbulence modelling of tokamak plasmas often exploits the axisymmetry of tokamak configurations to reduce the computational expense.  The nonaxisymmetric nature of stellarators, however, requires that simulations are fully three dimensional.  For three dimensional tokamak modelling, it is often advantageous to align the computational grid to the magnetic field, which helps improve numerical efficiency.  Typically, parallel dynamics exhibit a long wavelength, which allows for lower resolution in the parallel direction, and therefore faster computation. In stellarators, however, the complex magnetic geometry requires either a carefully designed field aligned system such as the unstructured grid used by the FINDIF code~\cite{Mctaggart2005}, or a nonaligned system since parallel nonuniformities are introduced along the magnetic field.

\boutxx~was originally developed for flute reduced plasma models in field aligned geometries.  Specifically, the three dimensions used were radial, toroidal, and parallel to the magnetic field.  This system is inaccurate in stochastic regions and regions near magnetic X- and O-points; two coordinates (toroidal, field aligned) are parallel at X-points, causing numerical instability in the form of zero-volume elements.  Recent work has sought to exploit techniques such as nonaligned coordinate systems which has allowed for X-point simulation in \boutxx~\cite{Shanahan2014,Shanahan2016,Leddy2016}.  We will show here that it is possible to simulate stellarator geometries in \boutxx~using a non-field-aligned grid through the implementation of the Flux Coordinate Independent (FCI) method for parallel derivatives.

\section{The Flux Coordinate Independent Method}
\label{sec:FCI}
 The advantage of field aligned coordinate systems is that parallel derivatives are simplified to be taken along one dimension of the coordinate system, which is computationally efficient since the majority of turbulence models are described by separating perpendicular and parallel derivatives.  One disadvantage of this method is that complex geometries such as magnetic nulls are poorly described and susceptible to numerical instabilities.  A second disadvantage of field-aligned coordinate systems is the difficulty associated with generating the mesh for nonaxisymmetric fields.  In most turbulence codes, field aligned grids are generated using a two dimensional poloidal cross section, and an assumption of axisymmetry which leads to a two dimensional equilibrium.  Additionally, the presence of magnetic islands and stochastic magnetic field regions in stellarators render this method impractical.  

\begin{figure}[htbp!]
\centering
\includegraphics[width=8cm]{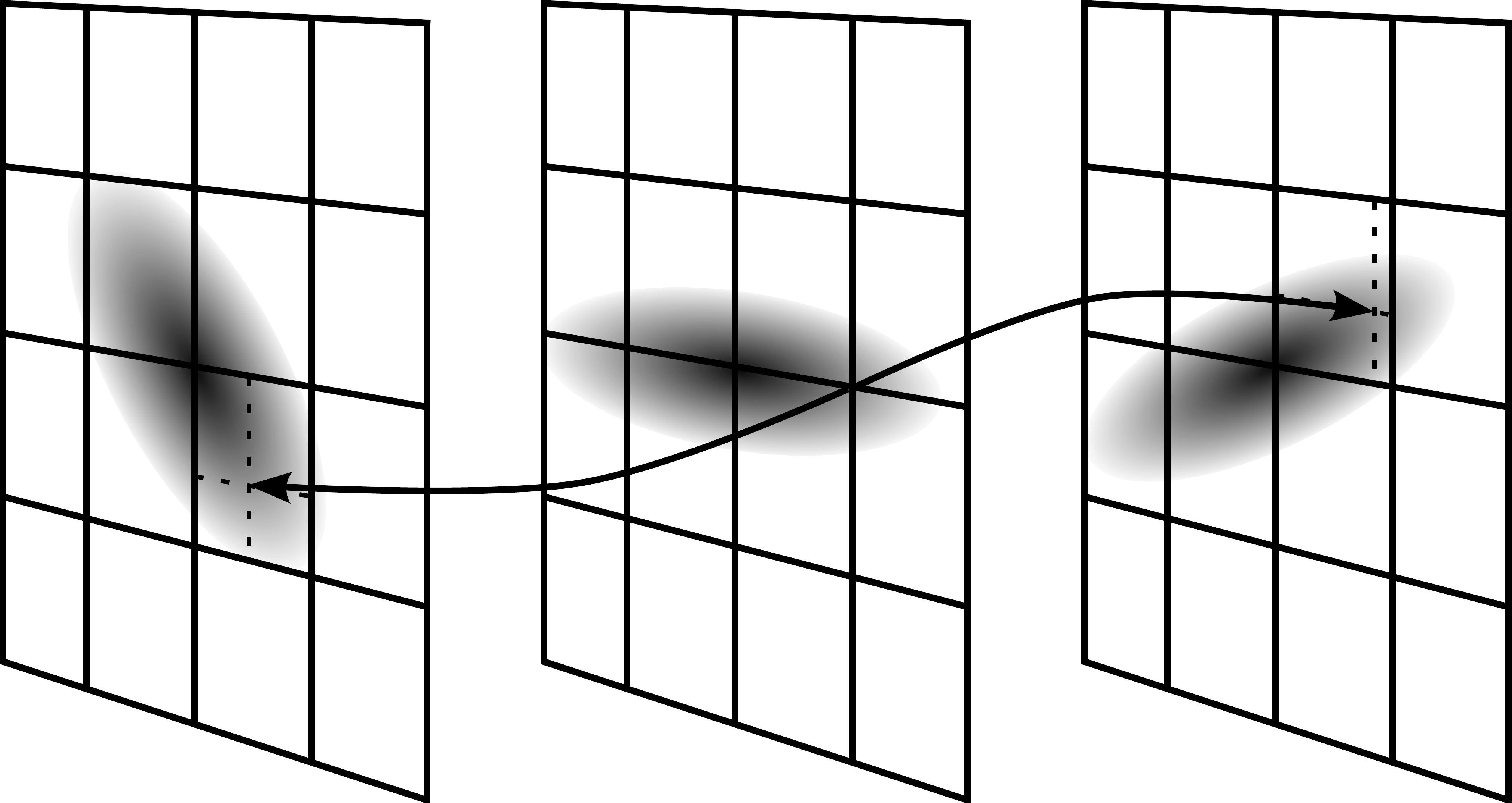}
\caption[Schematic of the FCI Method]{Illustration of Flux Coordinate Independent method for calculating parallel derivatives.}
\label{fig:FCI}
\end{figure}

Recently the Flux Coordinate Independent method for calculating parallel derivatives~\cite{Hill2016, Hariri2013} has been implemented in \boutxx.  This method for calculating parallel derivatives has been implemented in other codes~\cite{Stegmeir2016, Held2016}, and is intuitively straightforward, as described in Figure~\ref{fig:FCI}.

Based on the form of the magnetic field at a given point, the field line is followed to the next poloidal (or azimuthal) plane.  The position at which the field line hits the next plane is interpolated to the nearest grid points, and a value for a given quantity is assigned based on that interpolation. This process is repeated on the previous plane, and a differential is calculated based on central differencing. As the FCI method is used solely for calculation of parallel derivatives, the process is independent of the poloidal grid configuration.  In the work presented here, the poloidal grids are chosen to be Cartesian.

While this process is intuitively straightforward, it allows for more complex magnetic field configurations in comparison to structured grids. Previous work has used this model to simulate turbulence in the region of magnetic islands~\cite{Hill2015}, verifying its suitability for magnetic X- and O-points.  In the following section, the recent implementation of the FCI method is tested to determine if nonaxisymmetric modelling is possible within \boutxx.  

\section{Foundations for stellarator modelling in \boutxx}
\label{sec:boutstella}
The FCI method has been implemented into \boutxx~and tested via the method of manufactured solutions~\cite{Hill2016}, which determined that the operators converge to second order, as is expected for the central difference schemes in use.  One of the aims of the research presented here is to provide the components necessary to simulate stellarator turbulence cases and evaluate the computational and developmental work required.  The following subsection details the progress towards stellarator turbulence using the FCI method by describing the recent nonaxisymmetric test scenarios which have been implemented.

\subsection{Diffusion test cases}
\begin{figure}[h!]
\centering
\begin{minipage}{16pc}
\includegraphics[width=6.0cm]{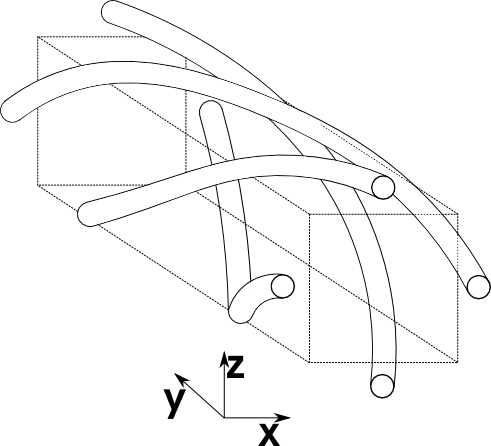}
\caption[The straight stellarator test case]{The straight stellarator test case; a very large aspect ratio classical stellarator showing helical coils and the inlaid Cartesian coordinate system (dashed).}
\label{fig:straightstell}
\end{minipage}\hspace{2pc}%
\begin{minipage}{16pc}
\includegraphics[scale=0.35]{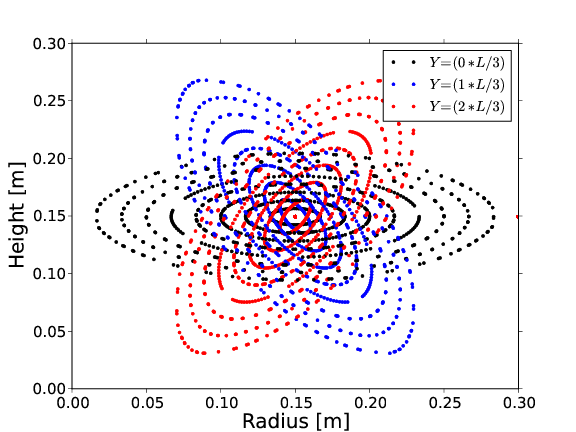}
\caption[Poincar\'e plot of the straight stellarator]{\label{fig:poincare} Poincar\'e plot of the straight stellarator indicating closed flux surfaces}
\end{minipage} 
\end{figure}

To determine the efficacy of the FCI method as a tool for stellarator turbulence modelling, a test case of an infinite aspect ratio classical stellarator was constructed.  A theoretical linear device with 4 helical coils was considered, as shown in Figure~\ref{fig:straightstell}.  A Poincar\'e plot of this configuration was created to ensure the existence of closed flux surfaces, as shown in Figure~\ref{fig:poincare}.

This configuration has been implemented into the FCI grid generator~\cite{Hill2016}, where it is possible to change several parameters including rotational transform, toroidal field and helical coil current/position. A typical grid with 16 Cartesian poloidal planes with a uniform 256 x 256 resolution can be generated in approximately 45 seconds.  

As the FCI method is purely a tool for calculating parallel dynamics, a simple heat advection equation was modelled:

\begin{equation}
\label{eq:diffusion}
\frac{\partial f}{\partial t} = \nabla \cdot \left({\bf{b}}{\bf{b}} \cdot \nabla f \right) \equiv \nabla^2_\parallel f
\end{equation}

\begin{figure}[h!]
\centering
\includegraphics[width=9.0cm]{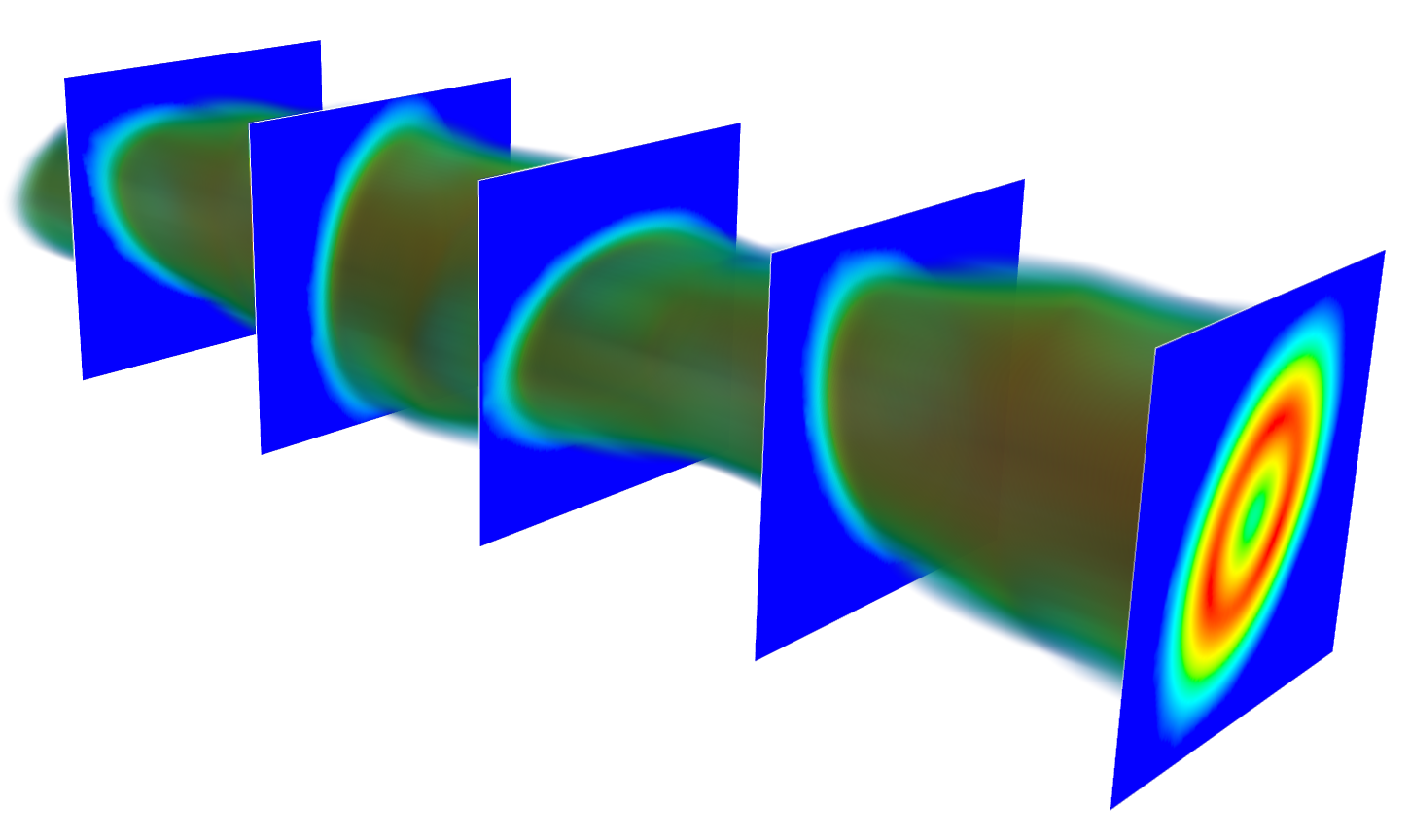}
\caption[Heat diffusion in the straight stellarator]{Heat diffusion in the straight stellarator test case; flux surfaces are correctly mapped out, qualitatively indicating proper calculation of parallel derivatives.}
\label{fig:diffusion}
\end{figure}

By solving Equation~\ref{eq:diffusion} for an initial three dimensional (non-field-aligned) Gaussian distribution of our test function $f$, and allowing the simulated to reach a saturated steady state, it is possible to trace out the flux surfaces (shown in Figure~\ref{fig:poincare}).  The results of the diffusion simulation are shown in Figure~\ref{fig:diffusion}.

Figure~\ref{fig:diffusion} indicates that the FCI method for calculating parallel derivatives is correctly evaluating parallel dynamics; as an initial distribution is left to propagate along the field lines, the flux surfaces are traced out.  The extent to which the interpolation of field lines in the perpendicular planes modifies the calculated quantities is tested in the following section.  

\subsection{Inherent Numerical Diffusion}
\label{sec:inherentnumericaldiffusion}
\begin{figure}[h]
\centering
\begin{minipage}{16pc}
\includegraphics[width=6.9cm]{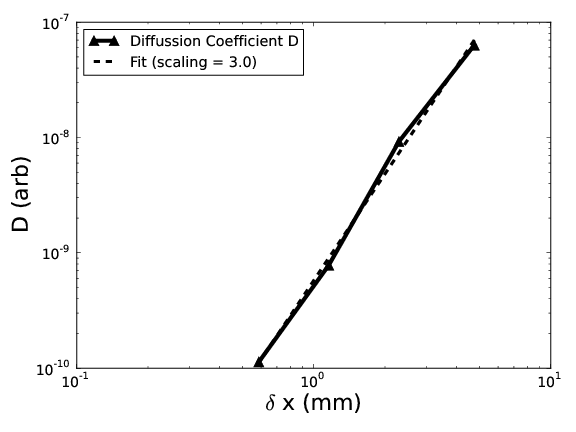}
\caption[Scaling of inherent diffusion from the FCI method in the straight stellarator]{Inherent numerical perpendicular diffusion as a function of poloidal mesh spacing in the straight stellarator test case.  The fit shows third order convergence, which is broadly in line with previous work~\cite{Hill2016}.}
\label{fig:diff_coeff}
\end{minipage}\hspace{2pc}%
\begin{minipage}{16pc}
\includegraphics[width=6.9cm]{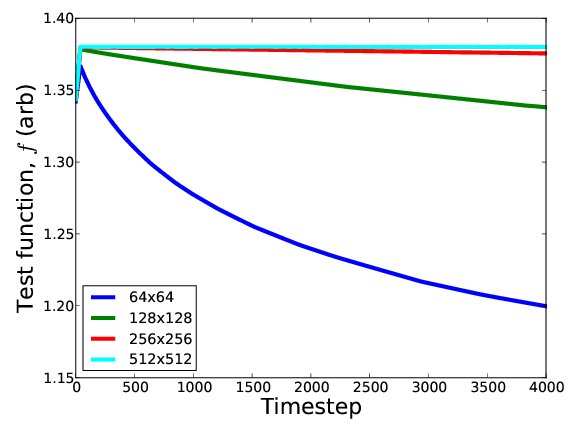}
\caption[Numerical diffusion as a function of time]{The test function $f$ at the center of the domain as a function of time for various resolutions; inherent numerical diffusion serves as an artificial sink at lower resolutions.}
\label{fig:nvt_FCI}
\end{minipage}
\end{figure}

There exists a strong anisotropy of heat conductivity in magnetized fusion plasmas.  Parallel conductivity can be as high as $\mathrm{10^{10}}$ times higher than perpendicular conductivity.  It is therefore important to minimize errors in parallel operators, as even a small perpendicular pollution of parallel dynamics can lead to substantial errors~\cite{Gunter2005}.  One of the main sources of error for the FCI method is the interpolation, as every quantity is interpolated based on where the field lines intersect the next and previous perpendicular planes.  The issue can be illustrated by considering a field aligned structure which is very small in the poloidal plane at a given grid point.  Assuming the field line does not intersect a grid point in the next (or previous) perpendicular plane, the structure will be distributed between the four nearest grid points.  This will then dissipate the function, causing a loss of accuracy and introducing an error.  Currently, the FCI method in \boutxx~utilizes cubic Hermite spline interpolation, but other methods are possible and can be used~\cite{Stegmeir2016}.  As a test of the interpolation, the previous diffusion case of a straight classical stellarator was implemented for 4 different grid resolutions.  It is possible to determine the inherent numerical diffusion from interpolation by assuming that the diffusion follows the relation:

\begin{equation}
\label{eq:D}
\frac{\partial f}{\partial t} = D \nabla^2_\perp f = D \nabla \cdot \left(\nabla f - {\bf{b}}{\bf{b}} \cdot \nabla f \right)
\end{equation}

where $D$ is the diffusion coefficient for the numerical diffusion of our test function.  
The inherent numerical diffusion from the interpolation scheme puts a limit on the minimum resolution which can be used, as low resolution grids will introduce a higher perpendicular numerical diffusion.  Ideally, numerical perpendicular diffusion should be at least a factor of $\mathrm{10^{-8}}$ smaller than the parallel dynamics~\cite{Gunter2005}.  The scaling of inherent perpendicular numerical diffusion coefficients with perpendicular mesh spacing in the straight stellarator geometry is shown in Figure~\ref{fig:diff_coeff}, where the diffusion coefficients at a grid point just off-axis (r,z = 16,15cm) are normalized to the parallel diffusion.  For this analysis, the number of parallel grid points was fixed at 16, and the number of perpendicular grid points varied; 64x64 resolution gives a mesh spacing of 4.76mm, 128x128 resolution gives a perpendicular mesh spacing of 2.29mm, 256x256 resolution indicates a mesh spacing of 1.16mm, and 512x512 resolution has a mesh spacing of 0.59mm.  For reference, the domain size is always set to 30cm x 3m x 30cm, and the parallel resolution is chosen to be a constant 19.6cm.  Figure~\ref{fig:nvt_FCI} indicates the loss of our test function $f$ due to numerical diffusion as a function of time for these various resolutions.  

From Figures~\ref{fig:diff_coeff} and~\ref{fig:nvt_FCI} it appears that the optimal resolution for an FCI mesh is 256 by 256, as the perpendicular diffusion is at least $\mathrm{10^{-8}}$ smaller than the parallel diffusion.  Of course, higher resolution cases are more precise but are also more computationally expensive.

Having quantified the diffusion inherent in the FCI method for parallel derivatives within \boutxx, we arrive at a minimum resolution required for this nonaxisymmetric configuration, which is comparable to the resolution one would choose for a given turbulence case for a system of this size, as $\rho_s \approx 1\mathrm{mm}$.  We have therefore provided evidence that nonaxisymmetric modelling is possible in \boutxx, as the most difficult barrier to stellarator modelling is the ability to correctly capture parallel dynamics. The next section describes recent work in implementing a transport model which is a subset of the EMC3 model, which intends to test the efficacy of \boutxx~as an alternative to common methods used in stellarator modelling.

\section{Transport modelling}
\label{sec:transportmodelling}
EMC3~\cite{Feng1999} is a commonly used tool to simulate the steady state profiles of stellarator edge plasmas, which allows for the reconstruction of heat flux profiles for divertor optimization. Recent work~\cite{Cosfeld2016} has looked to test the EMC3 model against analytic solutions of the one dimensional transport model.  Here, we present the first results following this work in \boutxx~to determine if \boutxx~can effectively and efficiently solve a simplified form of the EMC3 equations.  The EMC3 equations can be reduced to a one dimensional model which captures isothermal parallel dynamics, which are shown in~\cite{Cosfeld2016} to be:

\begin{equation}
\label{eq:EMC3n}
\frac{\partial n}{\partial t} = - \nabla \cdot \left( n v_\parallel\right)  + S_i
\end{equation}

\begin{equation}
\label{eq:EMC3nv}
\frac{\partial n v_\parallel}{\partial t} = - v_\parallel \nabla_\parallel nv_\parallel - 2 T_{e0} \nabla_\parallel n 
\end{equation}

where $T_{e0}$ is the isothermal electron temperature, assumed here to be 5eV, $n$ is the density, $S_i$ is the constant source function, and $v_\parallel$ is the parallel velocity. To ensure that \boutxx~is finding the correct solution, the simulated steady state of the above transport model compared with an analytical solution, which has been previously found to be~\cite{Cosfeld2016}:

\begin{equation}
\label{eq:EMC3n_ana}
n_i(x) = \frac{S_ix}{v_\parallel(x)c_s}
\end{equation}

\begin{equation}
\label{eq:EMC3nv_ana}
v_\parallel(x) = \frac{L}{2x} - \sqrt{\frac{L^2}{4x^2}-1}
\end{equation}

where $S_i$ is again the ion source which is independent of position, $x$ is the distance along the field line, $c_s$ is the sound speed and $L$ is the length of the domain, which spans from $-L$/2 to $L$/2.  For all of the following simulations, the boundary conditions were implemented in accordance with~\cite{Cosfeld2016}.  Specifically, velocities were set to $\pm c_s$ (which is normalized to 1) at the upper and lower boundary, respectively. Densities were set to $\frac{S_iL}{2c_s}$.

\begin{figure}[h!]
\begin{center}

\begin{minipage}{16pc}
\includegraphics[scale=0.35]{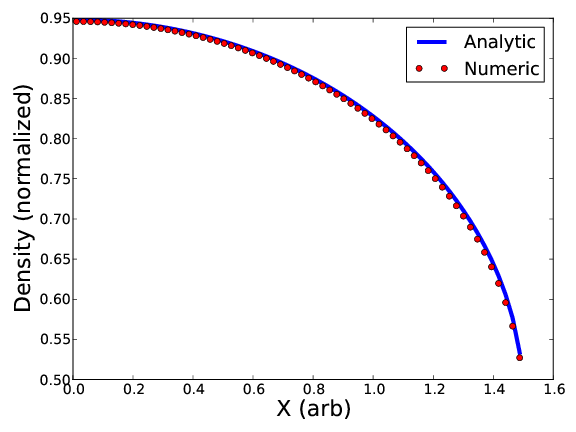}
\caption[Density in a 1D transport equation using finite volume operators]{\label{fig:EMC3n_FV} Steady state density in a 1D transport model and the analytical solution using conventional finite volume operators in \boutxx.}
\end{minipage}\hspace{2pc}%
\begin{minipage}{16pc}
\includegraphics[scale=0.35]{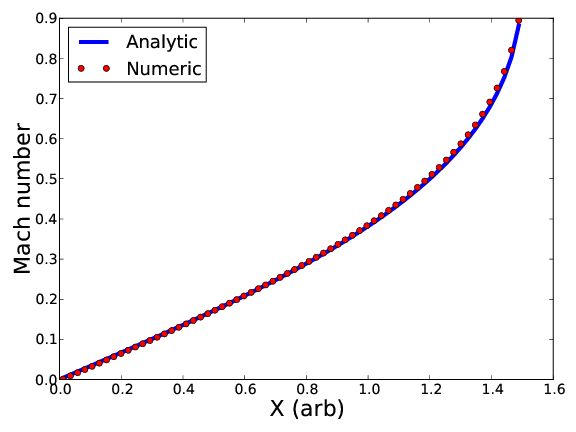}
\caption[Velocity in a 1D transport equation using finite volume operators]{\label{fig:EMC3v_FV} Comparison of the steady state velocity in a 1D transport model and an analytical solution using conventional finite volume operators in \boutxx.}
\end{minipage} 
\end{center}
\end{figure}

The first implementation of this model into \boutxx~was done without the use of the FCI method, allowing for testing using conventional operators.  Figures~\ref{fig:EMC3n_FV} and~\ref{fig:EMC3v_FV} illustrate a comparison of the analytical model and a simulation using finite volume (flux conserving) operators.  

\begin{figure}[h!]
\begin{center}

\begin{minipage}{16pc}
\includegraphics[scale=0.35]{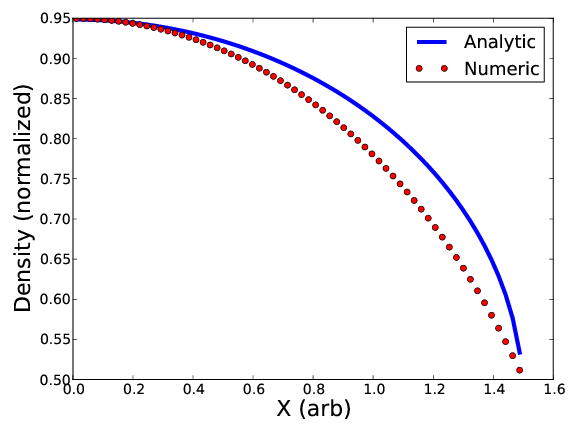}
\caption[Density transport solution simulated using finite difference operators]{\label{fig:EMC3n_FD} Steady state density in a 1D transport model and the analytical solution, having utilized a finite difference scheme within \boutxx.}
\end{minipage}\hspace{2pc}%
\begin{minipage}{16pc}
\includegraphics[scale=0.35]{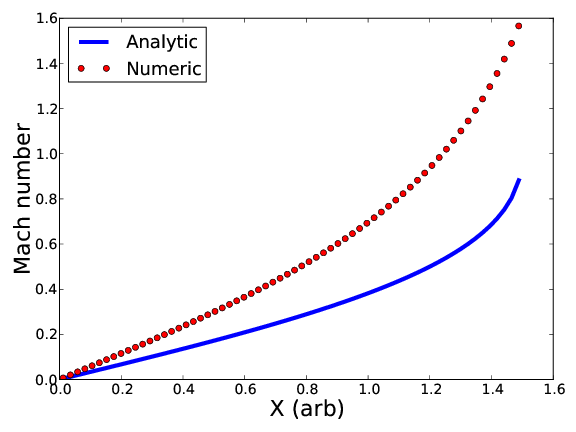}
\caption[Velocity solution using finite difference operators]{\label{fig:EMC3v_FD} Comparison of the steady state velocity using conventional finite difference schemes in a transport model and an analytical solution.}
\end{minipage} 
\end{center}
\end{figure}

It is clear from Figures~\ref{fig:EMC3n_FV} and~\ref{fig:EMC3v_FV} that \boutxx~is capable of simulating the correct profiles in these one dimensional transport equations.  However, the FCI method utilizes a finite difference scheme, which is not conservative and therefore could introduce losses.  As a test, the conventional (non-FCI) central differencing schemes were implemented and again compared to analytical solution, Figures~\ref{fig:EMC3n_FD} and~\ref{fig:EMC3v_FD}

Figures~\ref{fig:EMC3n_FD} and~\ref{fig:EMC3v_FD} indicate that the more simple central differencing scheme fails to reproduce the analytical solution, as the numerical steady state converges to incorrect profiles.  Using this method, it is possible that quantities can be lost from the simulation, as these operators are non-conservative; flux exiting one computational cell is not necessarily entering the next.  As such, it is important to minimize these losses when using a finite difference scheme, such as the FCI method. A method for improving these finite difference schemes will be discussed shortly.  

\begin{figure}[htbp!]
\centering
\includegraphics[width=8cm]{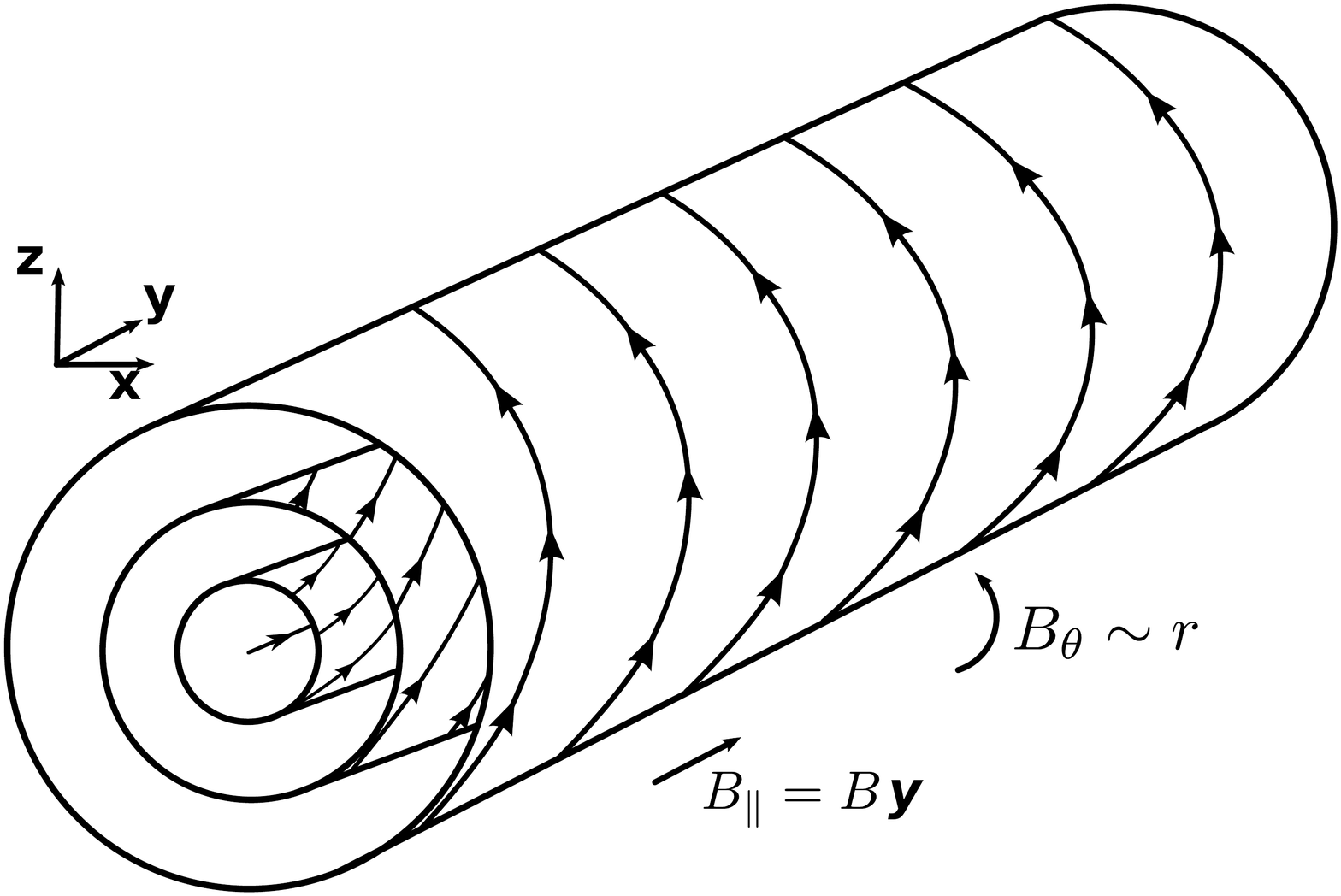}
\caption[Geometry for one dimensional FCI transport model test]{The geometry used to test the FCI method when solving the one dimensional transport model.  The azimuthal magnetic field is proportional to the minor radius $r$, allowing a test of straight field lines on axis, and a test of interpolation away from the center. }
\label{fig:EMC3_test}
\end{figure}

To further test the Flux Coordinate Independent method for parallel derivatives, this one dimensional transport model was implemented in the geometry shown in Figure~\ref{fig:EMC3_test}.  This geometry has a completely straight field at the center of the domain, where the FCI method must not interpolate, and an increasingly strong helical field at larger minor radii where interpolation is essential.  Furthermore, the magnetic field line length is nonuniform which (referring to Equation~\ref{eq:EMC3n_ana}) creates a radially varying profile for density. 

\begin{figure}[h!]
\begin{center}
\begin{minipage}{16pc}
\includegraphics[scale=0.35]{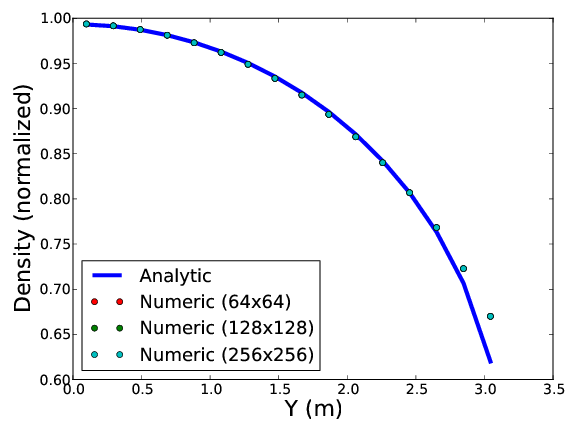}
\caption[The solution for density using modified finite difference operators]{\label{fig:EMC3n_FCI} Steady state solution for density in the one dimensional transport model and the analytical solution using the FCI method on straight field lines}
\end{minipage}\hspace{2pc}%
\begin{minipage}{16pc}
\includegraphics[scale=0.35]{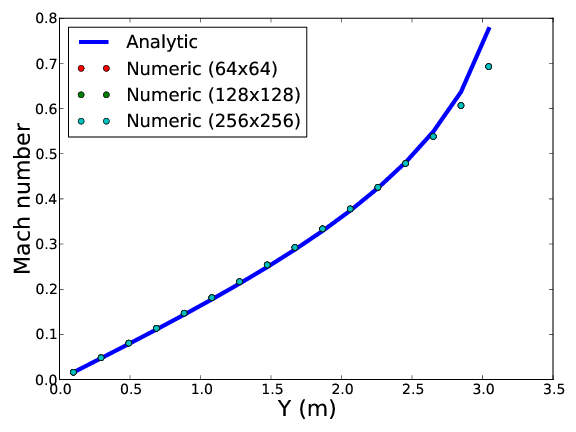}
\caption[The solution for mach number using modified finite difference operators]{\label{fig:EMC3v_FCI} Comparison of the steady state velocity and an analytical solution using the FCI method in a region of straight field lines.}
\end{minipage} 
\end{center}
\end{figure}

The one dimensional transport model was implemented into the geometry shown in Figure~\ref{fig:EMC3_test} at various resolutions.  Similar to the straight stellarator test case, the parallel resolution ($\textbf{y}$) was held constant while the resolution of Cartesian poloidal planes was varied. To minimize losses, the FCI operators were modified to calculate derivatives based on the flux at the at the grid cell faces; the flux at the edges of each computational cell was averaged with the flux at the edge of its neighboring cells.  While this still does not guarantee the conservation properties of finite volume methods, the quantities are more closely conserved than in simple finite difference schemes.  Figures~\ref{fig:EMC3n_FCI} and~\ref{fig:EMC3v_FCI} illustrate the results of the transport simulation at the center of the domain shown in Figure~\ref{fig:EMC3_test} for three different poloidal resolutions.  As these plots are taken at the center of the domain where the field line is straight, they can be compared to the previous finite difference results, Figures~\ref{fig:EMC3n_FD} and~\ref{fig:EMC3v_FD}, and a clear improvement is seen.

\begin{figure}[h!]
\begin{center}
\begin{minipage}{16pc}
\includegraphics[scale=0.35]{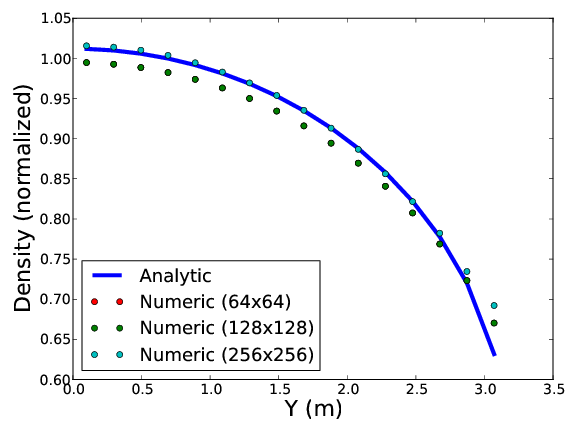}
\caption[The density solution for a transport model using FCI operators]{\label{fig:EMC3n_offcenter} Steady state density and the analytical solution of a one dimensional transport model in a region of helical field lines for three different resolutions.  Accuracy is increased with resolution.}
\end{minipage}\hspace{2pc}%
\begin{minipage}{16pc}
\includegraphics[scale=0.35]{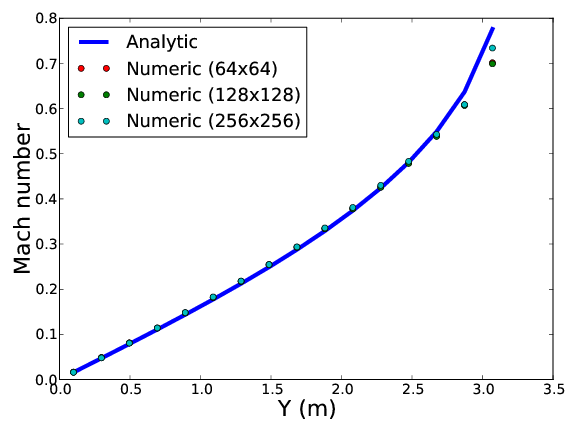}
\caption[The velocity solution for a transport model using FCI operators]{\label{fig:EMC3v_offcenter} Comparison of the steady state velocity and an analytical solution in a region of helical field lines.  Again, the higher resolution cases provide more accurate results.}
\end{minipage} 
\end{center}
\end{figure}

As the simulations correctly reproduces the behavior of the analytic model at the center of the domain, it can be concluded that the losses due to the central differencing scheme used by FCI have been reduced.  Furthermore, there is no dependence on poloidal resolution as the interpolation scheme is not used on straight field lines.  To test the interpolation of the FCI method, it is useful to examine the results away from the center where the field lines are helical, as shown in Figures~\ref{fig:EMC3n_offcenter} and~\ref{fig:EMC3v_offcenter}.  Specifically, these results were taken at about two-thirds of the distance to the edge of the computational domain, where the shear causes a shift of 2.35cm in the azimuthal direction between each perpendicular plane (separated by 19.6cm). 

Again, the simulations of the one dimensional transport model have reproduced the analytical solution along helical field lines, when considering a sufficiently high resolution (recalling Section~\ref{sec:inherentnumericaldiffusion}).  As these helical field lines require the use of interpolation in the FCI operators, these results indicate that errors due to interpolation are reduced to tolerable levels at sufficient resolution and the FCI method is capable of simulating transport models.

\section{Conclusions}
\label{sec:FCIconclusions}
Here we have discussed the recent progress in modelling nonaxisymmetric geometries within \boutxx.  The Flux Coordinate Independent approach to parallel derivatives has been implemented into \boutxx~and allows for complex geometries to be modelled.  A very large aspect ratio classical stellarator test case was implemented and it was determined that the FCI approach is correctly evaluating parallel dynamics, which was the most difficult challenge in modelling nonaxisymmetric configurations.  A one dimensional transport model was implemented and tested against an analytic solution, where it was determined that \boutxx~can effectively converge to the analytical solution using FCI operators.  These results indicate that \boutxx~has the components necessary to model nonaxisymmetric cases.  




\section{Acknowledgments}
This work has been carried out within the framework of the EUROfusion Consortium and has received funding from the Euratom research and training programme 2014-2018 under grant agreement No 633053. The views and opinions expressed herein do not necessarily reflect those of the European Commission. 

\section*{References}
\bibliography{varenna}
\bibliographystyle{unsrt}
\end{document}